\documentclass[nofootinbib,unsortedaddress]{revtex4-2}
\usepackage{amsmath}
\usepackage{dcolumn}
\usepackage{bm}
\usepackage{xcolor}
\usepackage{url}
\usepackage{graphicx}
\usepackage{float}
\usepackage{comment}

\usepackage[letterpaper,left=2.0cm,right=2.0cm,top=2.5cm,bottom=2.5cm]{geometry} 
\usepackage{siunitx} 
\usepackage{import} 
\usepackage{amssymb}
\usepackage{subcaption}

\usepackage[english]{babel}
\usepackage{hyperref}
\hypersetup{
    colorlinks=true,
    linkcolor=blue,
    filecolor=magenta,      
    urlcolor=blue,
    citecolor=blue,
}

\begin{document}

\title{Redshift Dipoles from Non-Geodesic Observer Congruences in Covariant Cosmology}

\author{Erick Past\'en}
\email{erick.contreras@usm.cl}
\affiliation{Departamento de F\'{\i}sica, Universidad de Santiago de Chile,
Avenida V\'{\i}ctor Jara 3493, Estaci\'on Central, 9170124, Santiago, Chile}

\begin{abstract}
Recent analyses of large-scale structure and redshift surveys have reported significant dipolar anisotropies in the local Universe that are not straightforwardly attributable to a kinematical boost. When interpreted within standard frameworks, these signals may correspond to coherent bulk flows that have been reported to exhibit tension with $\Lambda$CDM expectations. On the other hand, signals inferred from different astrophysical probes are not always consistent with the Cosmic Microwave Background (CMB) dipole, challenging the assumption of dipoles that are pure kinematical in origin.

In general cosmological scenarios, the congruence associated with cosmological observers need not be geodesic. Within a fully covariant framework, we show that a non-geodesic observer congruence introduces an additional contribution to the propagation of redshift along the past light cone, proportional to the line-of-sight projection of the observer four-acceleration. This generates an anisotropic modulation in the redshift itself, which could propagate to any observable defined in redshift space. We discuss different physical mechanisms that may produce a non-geodesic observer congruence, including interactions within the dark sector, modified theories of gravity, and relative motion between cosmological frames in tilted cosmologies.
\end{abstract}

\maketitle

\section{Introduction}

A growing amount of observational evidence suggests that the large-scale distribution of matter and the local expansion of the Universe may exhibit significant dipolar anisotropies. Recent analyses of redshift surveys and distance indicators have reported coherent large-scale dipoles in the Hubble flow and galaxy distribution that are not straightforwardly attributable to a kinematical boost \cite{Kalbouneh2026}. When interpreted within standard frameworks, these signals may correspond to coherent bulk flows whose amplitudes have been reported to exhibit tension with $\Lambda$CDM expectations \cite{Watkins2023}. More broadly, several large-scale dipolar or directional anomalies have been discussed in the literature, including the low-multipole alignments in the CMB and the unexpectedly large dipole reported in quasar number counts (see e.g. \cite{Rassat2014, Secrest2020, Dam2023}). Such observations illustrate that the interpretation of cosmic dipoles remains open and deserves careful scrutiny.

Cosmological observables are defined on the past light--cone of the observer, reflecting the causal structure of spacetime \cite{Ellis2012,KristianSachs1966}. Among them, redshift plays a central role, encoding the relation between spacetime dynamics and observational data. In standard cosmology, it is interpreted with respect to the comoving geodesic congruence, for which it reduces to the familiar relation $1+z = a_0/a$. However, this interpretation relies on a specific identification between the observer congruence and the underlying matter flow.

In this context, different proposed mechanisms may lead to non-geodesic large-scale flows or observer congruences with non-vanishing four-acceleration, including interacting dark-sector scenarios, modified theories of gravity, and relative motion between different cosmological frames in tilted cosmologies. In such scenarios, the interpretation of observables becomes particularly relevant, since the observer four-acceleration enters directly into the covariant propagation of photon energy along the past light cone and can imprint directional dependence on quantities inferred in redshift space \cite{Clarkson2012}. This raises the question about how different mechanisms that produce non-geodesic character in the selected cosmological congruence can affect the definition and interpretation of cosmological observables. In this work, we address this problem within a fully covariant framework \cite{Ellis2012,TSAGAS_2008}.

Observational effects induced by ``accelerated" observers are not merely theoretical. For example, the acceleration of the Solar System barycenter produces the well-known secular aberration drift \cite{Titov2011,Klioner2021}, a dipolar proper-motion pattern directly measured with very long baseline interferometry and Gaia astrometry. More generally, in fully general relativistic approaches, it has been shown that the 4-acceleration of the observer congruence induces a dipolar contribution to the redshifts of nearby sources ( see e.g. \cite{Clarkson1999}). In this case, the dipole arises as a local effect and grows with distance. This distinguishes it from a purely kinematic dipole, which is independent of distance and can be removed by a global boost. The mechanism considered here extends this picture to a fully covariant light-cone description, in which the relevant contribution is integrated along the line of sight. 

The structure of this paper is as follows. We first derive the Sachs optical equation for the angular diameter distance to emphasize its purely geometrical character. The propagation of redshift in both geodesic and non-geodesic observer congruences is then analyzed, showing that an extra anisotropic modulation naturally arises in the latter case. Finally, we discuss different physical mechanisms that may produce a non-geodesic observer congruence.

\section{The angular diameter distance from the Raychaudhuri equation}

Consider a congruence of null geodesics with tangent vector $k^\alpha$, affinely parametrised by $\lambda$,
\begin{equation}
k^\beta \nabla_\beta k^\alpha = 0,
\qquad
k^\alpha k_\alpha = 0 .
\end{equation}

Let $A(\lambda)$ denote the cross-sectional area of a narrow bundle of rays, defined on the two-dimensional screen space orthogonal to $k^\alpha$. The associated null expansion scalar is
\begin{equation}
\theta_N \equiv \frac{1}{A} \frac{dA}{d\lambda}.
\end{equation}

We define the optical shear and vorticity scalars of the null congruence as
\begin{equation}
\sigma_N^2 \equiv \frac12\,\sigma^{(N)}_{\alpha\beta}\sigma_{(N)}^{\alpha\beta} \ge 0,
\qquad
\omega_N^2 \equiv \frac12\,\omega^{(N)}_{\alpha\beta}\omega_{(N)}^{\alpha\beta} \ge 0,
\end{equation}
so that $\sigma^{(N)}_{\alpha\beta}\sigma_{(N)}^{\alpha\beta}=2\sigma_N^2$ and
$\omega^{(N)}_{\alpha\beta}\omega_{(N)}^{\alpha\beta}=2\omega_N^2$.

The evolution of the null expansion $\theta_N$ is governed by the null Raychaudhuri equation \cite{Raychaudhuri1955},
\begin{equation}
\frac{d\theta_N}{d\lambda}
=
-\frac12 \theta_N^2
- 2\sigma_N^2
+ 2\omega_N^2
- R_{\alpha\beta} k^\alpha k^\beta .
\end{equation}
With this convention, $\sigma_N^2$ contributes to focusing, while $\omega_N^2$ contributes to defocusing.

For a hypersurface-orthogonal null congruence (as appropriate for light bundles forming wavefronts), the vorticity vanishes,
\begin{equation}
\omega_N^2 = 0.
\end{equation}

Using $\theta_N=(1/A)\,dA/d\lambda$ and the identity (see Appendix \ref{app:identity_sqrtA})
\[
\frac{d\theta_N}{d\lambda}+\frac12\theta_N^2
=
\frac{2}{\sqrt{A}}\frac{d^2\sqrt{A}}{d\lambda^2},
\]
Raychaudhuri can be rewritten as an evolution equation for $\sqrt{A}$:
\begin{equation}
\frac{d^2 \sqrt{A}}{d\lambda^2}
=
-\frac12 \left(
R_{\alpha\beta} k^\alpha k^\beta + 2\sigma_N^2
\right) \sqrt{A}.
\end{equation}

For a shear-free light bundle,
\begin{equation}
\frac{d^2 \sqrt{A}}{d\lambda^2}
=
-\frac12 R_{\alpha\beta} k^\alpha k^\beta \sqrt{A}.
\end{equation}

The angular diameter distance is defined geometrically by
\begin{equation}
d_A^2 \equiv \frac{A}{\Omega},
\end{equation}
where $\Omega$ is the solid angle subtended by the source at the observer. The solid angle is defined in the observer's local rest frame at the vertex of the light cone and is therefore independent of the affine parameter along the null geodesic. Hence $d_A \propto \sqrt{A}$, and the evolution equation for $\sqrt{A}$ in the shear-free case reduces to
\begin{equation}
\frac{d^2 d_A}{d\lambda^2}
=
-\frac12 R_{\alpha\beta} k^\alpha k^\beta \, d_A .
\label{eq:sachs}
\end{equation}

Equation (\ref{eq:sachs}) corresponds to the shear-free limit of the Sachs optical equation \cite{Sachs:1961}, valid for hypersurface-orthogonal null congruences with vanishing vorticity. In this case, the evolution of the angular diameter distance is governed solely by the Ricci focusing term.

\subsection{Matter source term and FLRW universe}

Using Einstein's equations,
\begin{equation}
R_{\alpha\beta}
=
8\pi G \left( T_{\alpha\beta} - \frac12 T g_{\alpha\beta} \right)
+ \Lambda g_{\alpha\beta},
\end{equation}
and noting that $g_{\alpha\beta} k^\alpha k^\beta = 0$, we obtain
\begin{equation}
R_{\alpha\beta} k^\alpha k^\beta
=
8\pi G \, T_{\alpha\beta} k^\alpha k^\beta .
\end{equation}

For a perfect fluid,
\begin{equation}
T_{\alpha\beta}
=
(\rho + p) u_\alpha u_\beta + p g_{\alpha\beta}.
\end{equation}
Here $u^\alpha$ denotes the four-velocity of the cosmological perfect fluid sourcing the geometry. This implies
\begin{equation}
T_{\alpha\beta} k^\alpha k^\beta
=
(\rho + p) (u \cdot k)^2 .
\end{equation}

Thus the focusing term becomes
\begin{equation}
R_{\alpha\beta} k^\alpha k^\beta
=
8\pi G (\rho + p) (u \cdot k)^2 .
\end{equation}

Note that $-(u \cdot k)$ is the energy $E$ of the photon measured by the observer $u^\alpha$ and therefore
\begin{equation}
R_{\alpha\beta} k^\alpha k^\beta
=
8\pi G (\rho + p) E^2 .
\end{equation}

As a consistency check, specialising to an FLRW spacetime where the null congruence associated with radial light propagation is shear-free and the metric takes its standard form, the solution of Eq.~(\ref{eq:sachs}) is (see Appendix \ref{app:proof_SachsB})
\begin{equation}
d_A = a(\eta)\,S_K\!\big(\chi\big) .
\label{eq:dA_solution}
\end{equation}
with $\eta$ being the conformal time and $\chi$ the comoving distance.

\section{Redshift definition and the geodesic observer congruence}

From this point on, $u^\alpha$ will denote the observer congruence used to define measured photon energy and redshift; in exact FLRW it coincides with the fluid 4-velocity. The observed redshift is defined covariantly by
\begin{equation}
1+z_{obs} \equiv \frac{(u\cdot k)_{\rm em}}{(u\cdot k)_{\rm obs}},
\qquad
E \equiv -(u\cdot k),
\label{eq:z_def_nongeod}
\end{equation}

Differentiating $E=-(u\cdot k)$ along the ray gives
\begin{equation}
\frac{dE}{d\lambda}
=
-\,k^\alpha\nabla_\alpha(u_\beta k^\beta)
=
-\,k^\alpha k^\beta \nabla_\alpha u_\beta,
\label{eq:dE_general}
\end{equation}
where we used $k^\alpha\nabla_\alpha k^\beta=0$ (affine null geodesic). Introducing the standard 1+3 decomposition of $\nabla_\alpha u_\beta$:
\begin{equation}
\nabla_\alpha u_\beta
=
\frac13\theta\,h_{\alpha\beta}
+\sigma_{\alpha\beta}
+\omega_{\alpha\beta}
- u_\alpha \mathcal{A}_\beta,
\qquad
h_{\alpha\beta}=g_{\alpha\beta}+u_\alpha u_\beta .
\label{eq:1p3_decomp}
\end{equation}
For the photon, decompose $k^\alpha$ relative to the observers as
\begin{equation}
k^\alpha = E\,(u^\alpha+e^\alpha),
\qquad
u_\alpha e^\alpha=0,\qquad e_\alpha e^\alpha = 1,
\label{eq:k_decomp}
\end{equation}
where $e^\alpha$ is the spatial propagation direction measured by $u^\alpha$.

Substituting \eqref{eq:1p3_decomp} and \eqref{eq:k_decomp} into \eqref{eq:dE_general}, using $k^\alpha k_\alpha=0$ and $h_{\alpha\beta}u^\beta=0$, one obtains the exact kinematical propagation law:
\begin{equation}
\frac{1}{E}\frac{dE}{d\lambda}
=
-\,E\left(
\frac13\theta
+\sigma_{\alpha\beta}e^\alpha e^\beta
+ \mathcal{A}_\alpha e^\alpha
\right).
\label{eq:redshift_prop_general}
\end{equation}

For a geodesic observer congruence,
\begin{equation}
\mathcal{A}^\alpha \equiv u^\beta\nabla_\beta u^\alpha = 0.
\label{eq:geodesic_assumption}
\end{equation}

In the absence of shear in the observer congruence (as in exact FLRW), we have
\begin{equation}
\boxed{
\frac{1}{E}\frac{dE}{d\lambda}
=
-\,
\frac13 E \theta.
}
\label{eq:redshift_prop_geodesic}
\end{equation}

Equation \eqref{eq:redshift_prop_geodesic} reduces to the standard FLRW redshift law. Indeed, defining the expansion scalar and local Hubble rate of the \emph{observer congruence} by
\begin{equation}
H \equiv \frac{\theta}{3},
\label{eq:H_def_obs}
\end{equation}
one has
\begin{equation}
\frac{1}{E}\frac{dE}{d\lambda}=-EH.
\label{eq:Eprop_H}
\end{equation}
We define a redshift parameter along the null geodesic as $1+z(\lambda)=E(\lambda)/E_{\rm obs}$ along the ray whose value at the source position $\lambda=\lambda_s$ coincides with the observed redshift $z_{obs}$ and $E_{\rm obs}$ is constant at the observation event. It follows that
\begin{equation}
\frac{1}{1+z}\frac{d(1+z)}{d\lambda}
=
- EH.
\label{eq:zprop_H}
\end{equation}
On the other hand, in FLRW the scale factor depends only on cosmic time (see Appendix \ref{app:covariant_identities}), so
\begin{equation}
\nabla_\alpha a = -\dot a\,u_\alpha,
\qquad
\dot a = u^\beta\nabla_\beta a,
\label{eq:grad_a}
\end{equation}
Therefore,
\begin{equation}
\frac{da}{d\lambda}
=
k^\alpha\nabla_\alpha a
=
-\dot a\,u_\alpha k^\alpha
=
\dot a\,E
=
aHE.
\label{eq:da_dlambda}
\end{equation}
where we used $u_\alpha k^\alpha=-E$ and $H=\dot a/a$. Then
\begin{equation}
\frac{1}{a}\frac{da}{d\lambda}=EH.
\label{eq:aprop_H}
\end{equation}
Comparing \eqref{eq:zprop_H} and \eqref{eq:aprop_H}, we obtain
\begin{equation}
\frac{d\ln(1+z)}{d\lambda}
=
-\frac{d\ln a}{d\lambda},
\end{equation}

Integrating between the observer and source events yields
\begin{equation}
1+z_{obs}=\frac{a_0}{a},
\end{equation}
where $a$ is the scale factor evaluated at the source position. Thus, Eq.~\eqref{eq:redshift_prop_general} provides the exact covariant propagation law whose geodesic, shear-free FLRW limit reproduces the standard cosmological redshift relation.

\section{Non-geodesic observer congruence and the dipole modulation}
\label{subsec:nongeodesic}

The Sachs equation (\ref{eq:sachs}) for $d_A$ is purely geometrical and does not require the observer congruence to be geodesic. In the FLRW derivation above, the assumption of geodesicity entered when relating the affine parameter to the redshift parameter $z$ defined along the null geodesic. We now relax this assumption.

In the previous FLRW derivation we implicitly took $u^\alpha$ to be the comoving geodesic congruence. We now relax \eqref{eq:geodesic_assumption} and allow
\begin{equation}
\mathcal{A}^\alpha \neq 0.
\label{eq:nongeodesic_assumption}
\end{equation}

The explicit contribution of \eqref{eq:nongeodesic_assumption} to the redshift propagation is contained in the propagation of $E$ along the null ray. In full generality, the propagation law is
\begin{equation}
\boxed{
\frac{1}{E}\frac{dE}{d\lambda}
=
-\,E\left(
\frac13\theta
+\sigma_{\alpha\beta}e^\alpha e^\beta
+\mathcal{A}_\alpha e^\alpha
\right).
}
\label{eq:redshift_prop_full}
\end{equation}
The different kinematical contributions have a clear angular structure: the expansion scalar $\theta$ contributes to the monopole, the projection $\mathcal{A}_\alpha e^\alpha$ has a dipolar angular dependence, and the shear contribution $\sigma_{\alpha\beta}e^\alpha e^\beta$ has a quadrupolar structure. Equation \eqref{eq:redshift_prop_full} can be written as
\begin{equation}
\frac{1}{E}\frac{dE}{d\lambda}
=
- EH
- E\,\sigma_{\alpha\beta}e^\alpha e^\beta
- E\,\mathcal{A}_\alpha e^\alpha,
\end{equation}
where $H=\theta/3$. Using $1+z = E/E_{\rm obs}$, with $E_{\rm obs}$ constant, we obtain
\begin{equation}
\frac{1}{1+z}\frac{d(1+z)}{d\lambda}
=
- EH
- E\,\sigma_{\alpha\beta}e^\alpha e^\beta
- E\,\mathcal{A}_\alpha e^\alpha.
\label{eq:zprop_nongeod_full}
\end{equation}

In the present work, we focus on the direction-dependent modulation of
the redshift and therefore restrict attention to the anisotropic
contributions. We identify $a$ as an effective isotropic scale factor,
associated with the monopole of the expansion and defined
as
\begin{equation}
\frac{1}{a}\frac{da}{d\lambda}=EH.
\label{eq:effective_scale_factor}
\end{equation}
In the exact FLRW limit, this quantity coincides with the usual
cosmological scale factor. Equation~\eqref{eq:zprop_nongeod_full}
then integrates to 
\begin{equation}
(1+z_{\rm obs})
=
\frac{a_0}{a}
\exp\left[
-\int_{\lambda_0}^{\lambda_s}
E\left(
\mathcal{A}_\alpha e^\alpha
+\sigma_{\alpha\beta}e^\alpha e^\beta
\right)d\lambda
\right].
\label{eq:redshift_multipole_integrated}
\end{equation}

For weak departures from the geodesic FLRW case, this becomes
\begin{equation}
1+z_{\rm obs}
\simeq
\frac{a_0}{a}
\left[
1-
\int_{\lambda_0}^{\lambda_s}
E\left(
\mathcal{A}_\alpha e^\alpha
+\sigma_{\alpha\beta}e^\alpha e^\beta
\right)d\lambda
\right].
\label{eq:redshift_multipole_linear}
\end{equation}

This shows that non-geodesicity of the observer congruence induces an anisotropic modulation in the redshift with a local dipolar angular structure. Since the kinematical quantities may vary along the past light cone, these local angular structures do not necessarily translate into pure multipoles in the integrated redshift field. However, at low redshift a local expansion around the observer allows these quantities to be evaluated at leading order at the observer position, in which case the four-acceleration contribution reduces to a pure dipolar modulation \citep{Clarkson1999}.

\section{Physical mechanisms for a non-geodesic cosmological congruence}

In the following, we discuss some physically motivated mechanisms considered in the literature that can potentially produce a non-geodesic cosmological congruence.

\subsection{Momentum transfer in the dark sector}

A first interesting mechanism capable of generating a non-geodesic cosmological matter flow is the presence of momentum
exchange between different matter components \citep{Amendola2000,Valiviita2008,Pourtsidou2013,Linton2022}. In those scenarios, the energy-momentum tensors of dark matter and dark energy can satisfy
\begin{equation}
\nabla_\mu T^{\mu\nu}_{\rm DM}=Q^\nu_{\rm DM},
\qquad
\nabla_\mu T^{\mu\nu}_{\rm DE}=Q^\nu_{\rm DE},
\end{equation}
with
\begin{equation}
Q^\nu_{\rm DM}+Q^\nu_{\rm DE}=0,
\end{equation}
so that the total energy-momentum tensor remains conserved.
$Q^\nu$ is the interaction four-vector that can be decomposed with respect to the
dark-matter four-velocity $u^\mu$ as
\begin{equation}
Q^\mu_{\rm DM}
=
Q_{\rm DM} u^\mu
+
F^\mu_{\rm DM},
\qquad
u_\mu F^\mu_{\rm DM}=0,
\end{equation}
where $Q_{\rm DM}$ describes energy transfer in the dark-matter rest
frame and $F^\mu_{\rm DM}$ represents the corresponding momentum
transfer.

If we set the pressure of dark matter to zero we have
\begin{equation}
T^{\mu\nu}_{\rm DM}
=
\rho_{\rm DM}u^\mu u^\nu.
\end{equation}
The spatial projection of the conservation equation gives the Euler equation
\begin{equation}
\rho_{\rm DM}\mathcal{A}^\mu
=
F^\mu_{\rm DM}.
\label{eq:DM_euler_interaction}
\end{equation}
This shows that momentum transfer between dark matter and dark energy can make the dark-matter flow non-geodesic. This mechanism has been extensively discussed in the literature on interacting dark-sector models (see e.g. \cite{Koyama:2009gd}).

However, this mechanism cannot be directly connected with redshift anisotropies, since redshifts are measured from galaxies or supernovae that trace luminous baryonic matter. Therefore, a non-geodesic dark-matter flow does not directly imply that the baryonic congruence defining these observables is also non-geodesic. In this case, the interaction can generate relative motion between dark matter and baryons, but its connection with the four-acceleration entering the
redshift propagation law is not obvious. Nevertheless, interactions involving baryonic matter and dark energy have also been considered in the literature \citep{Vagnozzi:2019kvw, BeltranJimenez:2020iyx}. Such scenarios may give a more compelling observational relation with the
kinematical configuration considered in this work, since the
four-acceleration of the congruence traced by luminous matter can enter
the redshift propagation law directly.

\subsection{Some modified gravity scenarios}

Modified theories of gravity provide another possible mechanism for
generating non-geodesic matter flows. In this case, four-acceleration can emerge in theories involving non-minimal couplings between matter and curvature, producing an
additional force acting on matter. For example, an action of the form
\begin{equation}
S =
\int d^4x\,\sqrt{-g}
\left[
\frac{1}{2}f_1(R)
+
\left(1+\lambda f_2(R)\right)\mathcal{L}_m
\right]
\end{equation}
leads, in general, to a non-vanishing covariant divergence of the matter
energy-momentum tensor,
\begin{equation}
\nabla_\mu T^{\mu\nu}\neq0,
\end{equation}
as a consequence of the coupling between curvature and the
matter Lagrangian \citep{Bertolami:2007gv,Harko:2008qz}. As a result, the
equations of motion of matter contain an additional force orthogonal to the four-velocity, so that massive matter does not necessarily follow geodesics and the corresponding matter congruence can acquire
a non-zero four-acceleration.

\subsection{Tilted observer effects}

A conceptually different approach to a non-geodesic cosmological
congruence arises in tilted cosmologies, where two families of observers
move relative to each other. If $u^\mu$ denotes the background
congruence following the Hubble flow and $\tilde{u}^{\mu}$ denotes a
tilted congruence moving with a non-zero peculiar velocity relative to
the background, their four-velocities can be related by
\begin{equation}
\tilde{u}^{\mu}
=
\gamma\left(u^\mu+v^\mu\right),
\qquad
u_\mu v^\mu=0,
\end{equation}
where $v^\mu$ is the peculiar velocity field. At linear order around an FLRW background, their four-accelerations are related by
\begin{equation}
\tilde{\mathcal A}_\mu
=
\mathcal A_\mu
+
\dot v_\mu
+
H v_\mu ,
\end{equation}
showing that geodesicity in one frame does not, in general, imply
geodesicity in the other.

Currently, the physical interpretation of such an acceleration is subject to an active debate \citep{Tsagas:2026mgv,Clarkson:2026url} about whether the tilted four-acceleration should be regarded as a genuine physical effect. This question becomes particularly intricate when relating tilted observers to the large-scale matter flow, since reconciling the motion of individual observers with a smooth cosmological congruence description is itself non-trivial in an inhomogeneous Universe (see the discussion about peculiar velocities in \citep{Heinesen_2021}).

\section{Conclusions}

In this work, we have shown that the propagation of redshift along the past light cone depends explicitly on the kinematics of the observer congruence. In particular, relaxing the assumption of geodesicity introduces an additional contribution sourced by the projection of the observer four-acceleration, which induces an anisotropic modulation in the observed redshift. Unlike the standard kinematic dipole associated with a boost, this contribution is integrated along the photon trajectory.

This result has potential implications for the interpretation of large-scale dipoles in cosmological observations. Recent analyses of the local expansion rate based on CosmicFlows-4 and Pantheon+ data have reported a significant dipolar anisotropy \cite{Kalbouneh2026}. Within a standard framework, this signal can be interpreted as a large bulk flow, with amplitudes that are in tension with $\Lambda$CDM expectations \cite{Watkins2023}. This suggests either that peculiar velocities are larger than expected within standard cosmological scenarios \citep{Pasten:2026fwd,Tsagas:2026mgv}, or that the observed dipole may not admit a purely kinematic interpretation in terms of a single coherent bulk flow.

Our results show that such behaviour could, in principle, receive an additional contribution from non-geodesic effects in the redshift,
which may arise from several of the physical mechanisms discussed in this work. From an observational perspective, this opens the possibility of testing the geodesicity of the cosmological frame through the redshift and scale dependence of the observed dipole. Deviations from the redshift evolution expected from $\Lambda$CDM peculiar velocities, together with a correlation between the inferred dipolar signal and the predicted behaviour of a physically motivated congruence four-acceleration, could provide an observational signature consistent with non-geodesic effects.

\section*{Acknowledgements}

E.P. acknowledges support from the postdoctoral research project at Universidad de Santiago de Chile (USACH), project code 042531CM.

\appendix

\section{Evolution equation of $\sqrt{A}$}
\label{app:identity_sqrtA}

Let $A(\lambda)$ be the cross-sectional area of a narrow bundle of null rays, and let the null expansion be defined by
\begin{equation}
\theta_N \equiv \frac{1}{A}\frac{dA}{d\lambda}.
\label{eq:thetaN_def_app}
\end{equation}
We define
\begin{equation}
X(\lambda) \equiv \sqrt{A(\lambda)},
\qquad\Rightarrow\qquad
A = X^2 .
\label{eq:X_def_app}
\end{equation}
Differentiating with respect to $\lambda$ gives
\begin{equation}
\frac{dA}{d\lambda}=2X\frac{dX}{d\lambda}.
\label{eq:dA_app}
\end{equation}
Substituting \eqref{eq:dA_app} into \eqref{eq:thetaN_def_app} yields
\begin{equation}
\theta_N
=
\frac{1}{X^2}\left(2X\frac{dX}{d\lambda}\right)
=
2\,\frac{1}{X}\frac{dX}{d\lambda}.
\label{eq:thetaN_in_X_app}
\end{equation}
Now differentiate \eqref{eq:thetaN_in_X_app} once more:
\begin{align}
\frac{d\theta_N}{d\lambda}
&=
2\,\frac{d}{d\lambda}\left(\frac{1}{X}\frac{dX}{d\lambda}\right)\nonumber\\
&=
2\left[
\frac{1}{X}\frac{d^2X}{d\lambda^2}
-
\frac{1}{X^2}\left(\frac{dX}{d\lambda}\right)^2
\right].
\label{eq:dthetaN_app}
\end{align}
On the other hand, squaring \eqref{eq:thetaN_in_X_app} gives
\begin{equation}
\frac12\,\theta_N^2
=
\frac12\left(4\,\frac{1}{X^2}\left(\frac{dX}{d\lambda}\right)^2\right)
=
2\,\frac{1}{X^2}\left(\frac{dX}{d\lambda}\right)^2.
\label{eq:half_thetaN_sq_app}
\end{equation}
Adding \eqref{eq:dthetaN_app} and \eqref{eq:half_thetaN_sq_app}, the quadratic terms cancel identically, leaving
\begin{equation}
\frac{d\theta_N}{d\lambda}+\frac12\,\theta_N^2
=
2\,\frac{1}{X}\frac{d^2X}{d\lambda^2}.
\label{eq:identity_X_app}
\end{equation}
Finally, returning to $X=\sqrt{A}$, we obtain the useful identity
\begin{equation}
\boxed{
\frac{d\theta_N}{d\lambda}+\frac12\,\theta_N^2
=
\frac{2}{\sqrt{A}}\frac{d^2\sqrt{A}}{d\lambda^2}.
}
\label{eq:identity_sqrtA_app}
\end{equation}

\section{Proof that $d_A=aS_K(\chi)$ solves the Sachs equation in FLRW}
\label{app:proof_SachsB}

First note that along a radial null geodesic in conformal FLRW coordinates
\[
ds^2 = a^2(\eta)\left[-d\eta^2 + d\chi^2 + S_K^2(\chi)d\Omega^2\right],
\]
the null condition $ds^2=0$ for radial propagation implies
\begin{equation}
-d\eta^2 + d\chi^2 = 0
\qquad\Rightarrow\qquad
\frac{d\chi}{d\eta} = \pm 1.
\end{equation}
For an incoming ray propagating toward the observer, we take
\begin{equation}
\frac{d\chi}{d\eta} = -1.
\label{eq:dchi_deta}
\end{equation}

By definition of the tangent vector to the null geodesic,
\[
k^\eta = \frac{d\eta}{d\lambda},
\qquad
k^\chi = \frac{d\chi}{d\lambda}.
\]
Using \eqref{eq:dchi_deta}, we obtain
\begin{equation}
k^\chi
=
\frac{d\chi}{d\eta}\frac{d\eta}{d\lambda}
=
-\,k^\eta.
\label{eq:kchi_relation}
\end{equation}

Now consider an arbitrary scalar function $f(\eta,\chi)$.
The derivative along the null ray is
\begin{equation}
\frac{df}{d\lambda}
=
k^\alpha\nabla_\alpha f
=
k^\eta \partial_\eta f
+
k^\chi \partial_\chi f.
\end{equation}
Substituting \eqref{eq:kchi_relation}, we find
\begin{equation}
\frac{df}{d\lambda}
=
k^\eta\left(\partial_\eta-\partial_\chi\right)f.
\label{eq:d_dlambda}
\end{equation}

Applying this to $d_A=a(\eta)S_K(\chi)$ gives
\begin{equation}
\frac{d d_A}{d\lambda}
=
k^\eta\left(
\frac{da}{d\eta}S_K
-
a\frac{dS_K}{d\chi}
\right).
\end{equation}

Differentiating once more,
\begin{align}
\frac{d^2 d_A}{d\lambda^2}
&=
\left(\frac{dk^\eta}{d\lambda}\right)
\left(
\frac{da}{d\eta}S_K
-
a\frac{dS_K}{d\chi}
\right)
\nonumber\\
&\quad
+
(k^\eta)^2
\left(
\frac{d^2a}{d\eta^2}S_K
-
2\frac{da}{d\eta}\frac{dS_K}{d\chi}
+
a\frac{d^2S_K}{d\chi^2}
\right).
\label{eq:second_der_general_corrected}
\end{align}

To eliminate $dk^\eta/d\lambda$, we use the $\eta$-component of
the null geodesic equation,
\[
k^\beta\nabla_\beta k^\eta=0.
\]
In conformal FLRW coordinates, the relevant Christoffel symbols are
\[
\Gamma^\eta_{\eta\eta}
=
\Gamma^\eta_{\chi\chi}
=
\frac{1}{a}\frac{da}{d\eta}.
\]
Using $(k^\chi)^2=(k^\eta)^2$ for radial null propagation, we obtain
\begin{equation}
\frac{dk^\eta}{d\lambda}
+
\frac{2}{a}\frac{da}{d\eta}(k^\eta)^2
=
0,
\end{equation}
and therefore
\begin{equation}
\frac{dk^\eta}{d\lambda}
=
-\frac{2}{a}\frac{da}{d\eta}(k^\eta)^2.
\label{eq:geodesic_eta_corrected}
\end{equation}

Substituting this result into
\eqref{eq:second_der_general_corrected}, the terms proportional to
$dS_K/d\chi$ cancel identically, yielding
\begin{equation}
\frac{d^2 d_A}{d\lambda^2}
=
(k^\eta)^2
\left[
a \frac{d^2S_K}{d\chi^2}
+
\left(
\frac{d^2a}{d\eta^2}
-
2\frac{(da/d\eta)^2}{a}
\right)S_K
\right].
\label{eq:second_der_intermediate}
\end{equation}

Using
\begin{equation}
\frac{d^2 S_K}{d\chi^2}=-K S_K,
\label{eq:Sk_identity}
\end{equation}
we obtain
\begin{align}
\frac{d^2 d_A}{d\lambda^2}
&=
(k^\eta)^2 S_K
\left[
\frac{d^2a}{d\eta^2}
-
\frac{2}{a}\left(\frac{da}{d\eta}\right)^2
-
aK
\right]
\nonumber\\
&=
-(k^\eta)^2
\left[
2\left(\frac{1}{a}\frac{da}{d\eta}\right)^2
-
\frac{1}{a}\frac{d^2a}{d\eta^2}
+
K
\right]d_A .
\label{eq:second_der_final_conformal}
\end{align}

The first Friedmann equation in conformal time reads
\begin{equation}
\left(\frac{1}{a}\frac{da}{d\eta}\right)^2
+
K
=
\frac{8\pi G}{3}a^2\rho
+
\frac{\Lambda}{3}a^2.
\label{eq:Friedmann1_conformal}
\end{equation}
The corresponding acceleration equation is
\begin{equation}
\frac{1}{a}\frac{d^2a}{d\eta^2}
=
\frac{4\pi G}{3}a^2(\rho-3p)
+
\frac{2\Lambda}{3}a^2
-
K .
\label{eq:Friedmann2_conformal}
\end{equation}
Combining
\eqref{eq:Friedmann1_conformal} and
\eqref{eq:Friedmann2_conformal}, we obtain
\begin{equation}
\boxed{
2\left(\frac{1}{a}\frac{da}{d\eta}\right)^2
-
\frac{1}{a}\frac{d^2a}{d\eta^2}
+
K
=
4\pi G a^2(\rho+p).
}
\label{eq:conformal_identity}
\end{equation}

Substituting \eqref{eq:conformal_identity} into
\eqref{eq:second_der_final_conformal} gives
\begin{equation}
\frac{d^2 d_A}{d\lambda^2}
=
-4\pi G\,a^2(\rho+p)(k^\eta)^2\,d_A.
\label{eq:d2dA_final}
\end{equation}

For a perfect fluid,
\begin{equation}
R_{\alpha\beta}k^\alpha k^\beta
=
8\pi G(\rho+p)(u\cdot k)^2.
\end{equation}

For comoving observers in conformal FLRW coordinates,
\begin{equation}
u^\mu=\frac{1}{a}(1,0,0,0),
\end{equation}
and therefore
\begin{equation}
u\cdot k
=
g_{\eta\eta}u^\eta k^\eta
=
-a\,k^\eta.
\end{equation}
Equivalently, the photon energy measured by comoving observers is
\begin{equation}
E=-(u\cdot k)=a\,k^\eta.
\end{equation}

It follows that
\begin{equation}
R_{\alpha\beta}k^\alpha k^\beta
=
8\pi G(\rho+p)a^2(k^\eta)^2.
\label{eq:Rkk_FLRW}
\end{equation}

Comparing this expression with
Eq.~\eqref{eq:d2dA_final}, we immediately obtain
\begin{equation}
\frac{d^2 d_A}{d\lambda^2}
=
-\frac12 R_{\alpha\beta}k^\alpha k^\beta\,d_A,
\end{equation}
which is precisely the shear-free Sachs optical equation.

\section{Useful covariant identities}
\label{app:covariant_identities}

\subsection{Homogeneous scalars in FLRW}

In an exact FLRW spacetime, the scale factor $a$ depends only on cosmic time and is therefore homogeneous on the spatial hypersurfaces orthogonal to the comoving 4--velocity $u^\alpha$. This implies
\begin{equation}
h_\alpha{}^\beta \nabla_\beta a = 0,
\end{equation}
where $h_{\alpha\beta}=g_{\alpha\beta}+u_\alpha u_\beta$ is the spatial projector.

Using the standard 1+3 decomposition of the gradient of a scalar,
\begin{equation}
\nabla_\alpha a = -u_\alpha \dot a + h_\alpha{}^\beta \nabla_\beta a,
\qquad
\dot a = u^\beta \nabla_\beta a,
\end{equation}
it follows that
\begin{equation}
\nabla_\alpha a = -\dot a\,u_\alpha .
\label{eq:grad_a_app}
\end{equation}

\subsection{Contractions involving the photon wavevector}

The photon wavevector is decomposed relative to the observer congruence as
\begin{equation}
k^\alpha = E(u^\alpha+e^\alpha),
\qquad
u_\alpha e^\alpha=0,
\qquad
e_\alpha e^\alpha=1,
\qquad
u_\alpha u^\alpha=-1.
\label{eq:k_decomp_app}
\end{equation}
From this decomposition one obtains the following useful identities.

First,
\begin{equation}
k^\alpha k^\beta h_{\alpha\beta}
=
E^2 (u^\alpha+e^\alpha)(u^\beta+e^\beta)h_{\alpha\beta}
=
E^2 .
\label{eq:khh_app}
\end{equation}

Second, since the shear tensor is spatial, symmetric and orthogonal to $u^\alpha$,
\begin{equation}
k^\alpha k^\beta \sigma_{\alpha\beta}
=
E^2 \sigma_{\alpha\beta}e^\alpha e^\beta .
\label{eq:ksigma_app}
\end{equation}

Third, since the vorticity tensor is antisymmetric,
\begin{equation}
k^\alpha k^\beta \omega_{\alpha\beta}=0.
\label{eq:komega_app}
\end{equation}

Finally, using $u_\alpha k^\alpha=-E$ and $u_\alpha \mathcal{A}^\alpha=0$, one has
\begin{equation}
k^\alpha k^\beta u_\alpha \mathcal{A}_\beta
=
E^2 (u^\alpha+e^\alpha)(u^\beta+e^\beta)u_\alpha \mathcal{A}_\beta
=
- E^2\, \mathcal{A}_\alpha e^\alpha .
\label{eq:kA_app}
\end{equation}

\bibliographystyle{JHEP}
\bibliography{mybib}

\end{document}